\documentclass{article}  

\usepackage[sort&compress]{natbib}
\usepackage[pdftex]{graphicx}
\usepackage[nolist]{acronym}
\usepackage{algorithm}
\usepackage{algpseudocode}
\usepackage{url}

\newcommand{\addwidefig}[4]
{
\begin{figure}[tp]
\centerline{\includegraphics[width=200pt,angle=0,trim=110 70 350 250]{#2#3}}  
\caption{#4}
\label{fig:#1}
\end{figure}
}

\date{}
\author{J. Wiesinger, D. Sornette, J. Satinover}
\title{Reverse Engineering Financial Markets with Majority and Minority Games using Genetic Algorithms}

\begin{document}

\begin{acronym}[delGCMjG]  
 \acro{EMH} {Efficient Market Hypothesis}
 \acro{ABM} {agent-based model}
 \acro{GA} {genetic algorithm}
 \acro{SGA} {Simple Genetic Algorithm}
 \acro{3PG} {Third Party Game}
 \acro{MG} {Minority Game}
 \acro{GCMG} {``Grand Canonical" Minority Game}
 \acro{GCMjG} {``Grand ``Canonical" Majority Game}
 \acro{delGCMG} {Delayed ``Grand Canonical" Minority Game}
 \acro{delGCMjG} {Delayed ``Grand Canonical" Majority Game}
 \acro{MixG} {Mixed Game}
 \acro{ISD} {initial strategy distribution}
\end{acronym}

\maketitle 

\begin{abstract}

Using virtual stock markets with artificial interacting software investors, aka \acp{ABM}, we present a method to reverse engineer real-world financial time series. 
We model financial markets as made of a large number of interacting boundedly rational agents.  
By optimizing the similarity between the actual data and that generated by the reconstructed virtual stock market, we obtain parameters and strategies, which reveal some of the inner workings of the target stock market.
We validate our approach by out-of-sample predictions of directional moves of the Nasdaq Composite Index.

\end{abstract}

``What I cannot create, I cannot understand'': On physicist Richard Feynman's blackboard at time of death in 1988; as quoted in The Universe in a Nutshell by Stephen Hawking.

\section{Introduction}

The prediction of financial markets has long been the object of keen interest among both financial professionals and academics. The widely, - if not universally -, accepted  \ac{EMH} \citep*{Fama1970Efficient}, \citep*{Fama1991Efficient} provides a powerful argument that markets are inherently unpredictable, in particular on the basis of prior price data: Because all information about the future is incorporated into the current price (for all practical purposes immediately), price changes must follow a random walk \citep*{Malkiel2003Random}. There is considerable evidence however that prices do not perfectly follow a random walk and that some price inefficiency is present, varying over time, perhaps enough at times to be exploitable \citep*{Dahlquist1998Technical}. However, recent assessments of the performance of hedge-funds \citep*{Barras2008False} and of mutual funds \citep*{Fama2009Luck} cast doubt on the 
reality of the gains resulting from the practical implementation of these inefficiencies, if they exist.
As illustrated in the approaches of \citet*{Barras2008False} and \citet*{Fama2009Luck},
deviations from the EMH are searched in the form of anomalous performance, beyond what can be explained by risk premia associated with exposures to a few dominating risk factors. 

The near-absence of predictability in financial markets, or more precisely of risk-adjusted arbitrage opportunities,  is truly remarkable.
A rich academic literature has clarified the zen-like nature of the \ac{EMH} in the
sense that, the more intelligent are the investors and the harder are their efforts
to gather information to make the best possible investment decisions, the fewer trading opportunities there are, and the more efficient is the market.
The fact that markets are close to efficient can thus be understood as a macroscopic
organization that result from the collective actions of the active investors. 
Borrowing from the jargon of complex system theory, market efficiency is an emergent phenomenon. 
Emergence, the existence of qualitatively new properties exhibited by collections of interacting individuals, is often taken to be the defining characteristic of complex adaptive systems. 

Reciprocally, we ask here
how the observation of the large scale behavior of a macroscopic system can (i) uncover the internal properties of a system and the organization among its constituents and (ii) be used for its prediction.
Following Richard Feynman, we argue that, in order to really understand
a system, we need to be able to strip things down, then rebuild them in order to play with the reconstructed simplified system and analyze variants, from which understanding can emerge.
We address this question of ``reverse engineering'' in the context of one-dimensional financial (market) time-series. The challenge consists in building a virtual stock market with artificial interacting software investors.
The method presumes that real-world discrete market price changes may be in principle modeled as the aggregated output of a large number of interacting boundedly rational agents. These agents have limited knowledge of the detailed properties of the markets they participate in and create, have access to a finite set of strategies to take only a small
number of actions at each time-step and have restricted adaptation abilities.
Given the time series data, our method of reverse engineering determines
what set of agents, with which parameters and strategies, optimizes (in the sense of 
various robust metrics) the similarity between the actual data and that
generated by an ensemble of virtual stock markets peopled by software investors.
We provide a validation step by testing the performance of the reverse engineered artificial market in predicting out-of-sample directional moves of the real-world time series. Using only some of the simplest strategies and agents, the $p$-value for the statistical significance of the prediction of the directional moves for more than 600 trading days of the Nasdaq Composite Index is smaller than $0.02$. The results are robust with changes of the styles of agents' strategies and for different market regimes.

Our work uses the extensive literature on agent-based models that has been developing at
least since the 1960s (see \citet*{LeBaron2000Agentbased} and references therein). 
In \acp{ABM}, a system is modeled as a collection of autonomous decision-making entities, called agents. Repetitive competitive interactions among agents generate complex behavioral patterns. Due to the evolutionary switching among strategies, \acp{ABM} are highly nonlinear. The aggregation of simple interactions at the micro level may generate sophisticated structures at the macro level  which provide valuable information about the dynamics of the real-world system which the \ac{ABM} emulates. The main benefits of \acp{ABM} are that they (i) capture emergent phenomena; (ii) provide a natural description of a system; (iii) are flexible. \acp{ABM} have already been successfully applied in real-world problems, such as, flow simulation, organizational simulation, diffusion simulation and market simulation \citep*{Bonabeau2002Agentbased}. In this article we focus on financial market simulation. 

\citet*{Hommes2006Chapter, Hommes2002Modeling} shows that \acp{ABM} can explain the main statistical regularities observed in financial time series - their so-called ``stylized facts'' - such as excess volatility and volatility clustering, high trading volume, temporary bubbles and trend following, sudden crashes and mean reversion, and fat tails in the distribution of returns. Toy models such as the \ac{MG}, described in detail in \citet*{Challet1997Emergence}, capture key features of one generic market mechanism (competition for a scarce resource). The basic interaction between agents and public information is described in \citet*{Challet2001From, Marsili2001Market}.  Details of the ABMs we employ will be introduced as we describe the implementation of our 
reverse engineering process. In brief, we concentrate on the so-called \ac{MG} and its key variants and on the so-called \$-game and related majority games.

A major thrust of the literature of \acp{ABM} dealing with finance is aimed at developing artificial stock markets and then analyzing the conditions which yield the stylized facts of real markets. Changes of parameters or of the model proper affect the collective behavior of the model and thus provide potential insight into the underlying structure of the real-world market. We take this one step further and focus on reverse engineering specific financial markets with the help of \acp{ABM}. Reverse engineering means that we are trying to find a generating process of a real financial time series based on the time series itself. We provide a first validation step, not by quantifying how well the reconstructed synthetic market explains stylized facts but, rather by testing simple predictability. 

In \citet*{Jefferies2000From, Johnson2001Application}, the authors developed a first reverse engineering approach, using a \ac{GCMG}, whose detailed description is found in \citet*{Johnson2000Trader}. The \ac{GCMG} is an extension to the basic \ac{MG}, in which the total number of actively participating agents fluctuates. The authors did not report results using real financial time series, but a time series generated by a known ensemble of such agents they pretend to know nothing about this ensemble apart from its output.
Hence, they called it a ``black box" ensemble. They then began with an ensemble of agents with randomized parameters (so-called \acp{3PG}) and, by iteration, ``evolve" in parameter space this ensemble of \acp{3PG} until its output matches that of the known, black box, ensemble. Here, matching meant maximizing the cross-correlation of the black box and \ac{3PG} time series. One may then open the black box to determine how well this procedure has approximated the structure of the unknown black box. When the evolutionary process is successful, this can be applied to a real world series. In the sequel, we will follow this general procedure and treat heuristically the resulting \ac{3PG} ensemble as a model of the truly unknown real world market structure of traders.

The main challenge in this procedure is finding an adequately optimized set of parameters for the \ac{3PG}, as parameter space is large and grows extremely rapidly with every increasing level of sophistication. Furthermore, the landscape of the solution space is extremely rugged, reflecting the underlying degree of frustration among competing agents in the model (and presumably, in the market being modeled). For this search, we use a \ac{GA}, which is a methodology that adopts evolution used in nature to optimize the adaptation of life to the environment \citep*{Holland1992Adaptation, Goldberg1989Genetic}. For example in \citet*{Lettau1997Explaining, Arifovic1996Behavior, Chen2008Prediction, Palmer1994Artificial}, \acp{GA} are successfully used to equip agents with learning behavior for acting more profitably. 
We apply a \ac{GA} not for the individual learning process of the agents but for finding an ensemble of agents and their strategies best able to reproduce the time series that we hope to predict, referred to as the external time series throughout the article. In \citet*{Andersen2005Mechanism}, a prototype is developed which identified a new mechanism for short-term predictability in \acp{ABM}.
In order to test the validity of this approach, i.e., to test how well the generating process of the time series can be captured by the reverse engineered \ac{ABM} \ac{3PG}, we analyze the predictions we obtain from the identified \ac{3PG} when it must predict  out-of-sample real financial data.

\section{Model / Methodology}
\label{sec:model}

\subsection{General set-up of the reverse engineering method}

\begin{figure}[tp]
\centerline{\includegraphics[width=280pt,angle=0]{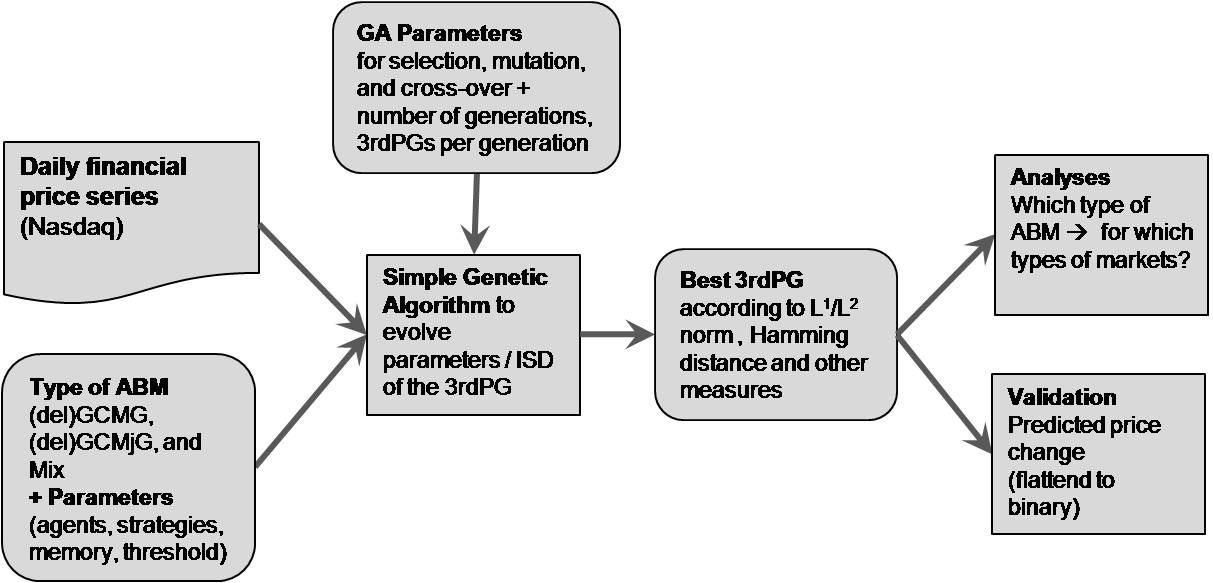}}
\caption{demonstrates a process overview of how a time series of daily adjusted closing prices is fed to a \ac{GA} along with different parameters determining (i) the type of \ac{ABM} for the \ac{3PG} which is used for generating a similar time series to the real one (during the in-sample period) and (ii) the convergence behavior of the \ac{GA} for the search of the best - most similar - \ac{3PG} according to measurements like $L^{1}$, $L^{2}$ and the Hamming distance. This result is then analyzed with respect to the types of \acp{ABM} present in which types of markets. The generating process is validated by the accuracy of one-step predictions.}
\label{fig:globalPicture}
\end{figure}


Figure \ref{fig:globalPicture} illustrates the whole process from the input to the prediction which will be explained stepwise in the following. In a nutshell, given a financial time series over some time interval and for a fixed  
\ac{ABM}, using a \ac{GA} (specified by its structure and parameters governing its search), we select a set of ``best" \acp{ABM} i.e., their output best matches the financial time series. By ``best" match we mean a minimization of ``distances'' between the financial time and the \ac{3PG} series, based on correlations \citep*{Lamper2001Predictability} and different standard norms. The results are found robust with respect to the choice of these norms.

\subsection{The Nasdaq Composite index as the input of the reverse engineering process}

As input to our model, we use daily adjusted closing prices of the Nasdaq Composite index. 
We assess the performance of the reverse engineering approach from its ability to predict the signs of out-of-sample returns on the same Nasdaq time series.
Results are obtained for 606 predictions.
We present both aggregated metrics as well as results sorted according to 
different market regimes (upward trend, downward trend and no-trend), and compare
with standard benchmarks i.e., with buy-and-hold (winning in upward trends), sell-and-hold (winning in downward trends), and random strategies. By distinguishing the three market regimes, we can infer from the performance of different type of \acp{ABM} which population of investors were dominant. For instance, it is intuitive and we confirm that trend-following strategies are dominant during upward trending markets. More surprising is the evidence we find for contrarian (or minority-type) strategies also performing well during such market phases, as we describe below.

The size of our statistical tests over 606 predictions constitutes a significant improvement with respect to prior effort of \citet*{Andersen2005Mechanism}, which
dealt with only a few tens of predictions. We were able to improve on this previous work
using more efficient coding and the access to more computer resources available at ETH Zurich through the Brutus super-cluster. While much larger, our sample size remains  limited by the high computational processing costs associated with the search of the \ac{GA} exploring a large parameter space.

\subsection{Description of the different types of \acp{ABM}}
\label{sec:abmTypes}

While we use different \acp{ABM} to be described shortly, 
the following properties are common to all of them. For a given \ac{ABM} with $N$ agents, each agent has to repeatedly choose among buying, selling or staying out of the market, according to their strategies. The agents base their decision on (i) the previous performance of their strategies indicated by the virtual point counters, (ii) their threshold - is it profitable to trade with their strategies? - and (iii) their memory of prior returns - in its binary representation (up / down) - of the external time series. 

The following types of \acp{ABM} are used, which differ in the incentives provided to the agents.
\begin{enumerate}
	\item \textbf{\acf{GCMG}}. In the \ac{GCMG}, an agent is rewarded for being in the minority \citep*{Johnson2000Trader}, whereby the extension of the classical \ac{MG} consists therein that an agent has the possibility not to trade and hence, allowing for a fluctuating number of agents invested in the stock market.
	
	\item \textbf{\acf{GCMjG}}. In the GCMjG, an agent is rewarded for being in the majority instead of in the minority \citep*{Marsili2001Market}.
	
	\item \textbf{\acf{delGCMjG}}. In the \ac{delGCMjG}, an agent is rewarded similarly to an agent in the \ac{GCMjG}
	but for the fact that the return following the decision is delayed by one time step, in order to reflect the more realistic market property that returns are accrued after some time following an investment decision.
	The grand canonical version is derived from the so-called \$-game introduced by  \citet*{Andersen2003Game}.
		
	\item \textbf{\acf{delGCMG}}. This game is the analog of the \ac{delGCMjG}, except for the minority payoff, whereby each agent is rewarded according to how the return at the next time step is compared with her decision taken at the previous time step. In other words, the delGCMG is a delayed \ac{GCMG}.
	
	\item \textbf{\acf{MixG}}. In the version of the MixG used here,  we consider a mix of agents, with 50\% of the agents obeying the rules of the \ac{GCMG} and the other 50\% obeying the rules of the \ac{GCMjG}. 
\end{enumerate}

\subsection{Description of the \acl{GA}}

The \ac{3PG} which best reproduces the external time series provides the solution
to our reverse engineering problem.  This \ac{3PG} is determined from a search
in the space of parameters of the ABM using a \ac{SGA} as shown in Algorithm \ref{alg:SGA}.
First a population of \acp{3PG} is initialized, whereby the number of agents, the number of strategies an agent obtains, the size of her memory, and her threshold are constant in the current version of the \ac{SGA}. The only aspect in which the \acp{3PG} differ is the \ac{ISD} which is the crucial parameter set over which we optimize the fit to an external time series.

\begin{algorithm}
	\footnotesize{\begin{algorithmic}
	  \Function{\acs{SGA}}{$extReturns, fitness(\cdot)$}
	    \State $t\gets$ 0  \Comment{Time in nbr of generations}
	    \State $p\gets p_0$   \Comment{Initialization of \ac{3PG}s}
	    \While{(not terminal condition)}   \Comment{Evolution}
	    	\State $t\gets t$+1
	      \State $fitness(p_{t-1}, extReturns)$  \Comment{Calculate the fitness}
	      \State $p_t \gets crossOver( selection(p_{t-1}) )$  \Comment{Create a new generation}
	      \State $mutation(p_{t})$  \Comment{Mutate randomly}
	    \EndWhile
	    \State\Return $bestOf(p_{t})$  \Comment{Return best \ac{3PG}}
	  \EndFunction
	\end{algorithmic}}
\caption{\acl{SGA}.}
\label{alg:SGA}
\end{algorithm}

For the first generation, the \acp{ISD} are initialized randomly. Then for every \ac{3PG}, its fitness - reflecting how well the time series generated by the \ac{3PG} matches the external time series - is determined. This value is computed via a fitness function using different metrics, such as the $L^{1}$ and $L^{2}$-norms, the 
Hamming distances (with binary and ternary coding) between the two time series. 

\acp{3PG} are selected to produce offspring according to their fitness, with the fittest yielding more offspring.
Each new generation of \acp{3PG} is obtained as a mixture of the agents and of the strategies of the previous parent generation. Many generations evolve until a convergence criterion is reached, which leads us
to finally identify a \ac{3PG} which best represents the external time series within the in-sample period.
The search is performed ten times to obtain ten \acp{3PG}. The differences between these ten solutions provide a measure of the quality of the reverse engineering method. The ten \acp{3PG} are also used to quantify the uncertainty in the next-day out-of-sample prediction.
 
\addwidefig{3rdPGvsExtReturn}
{}{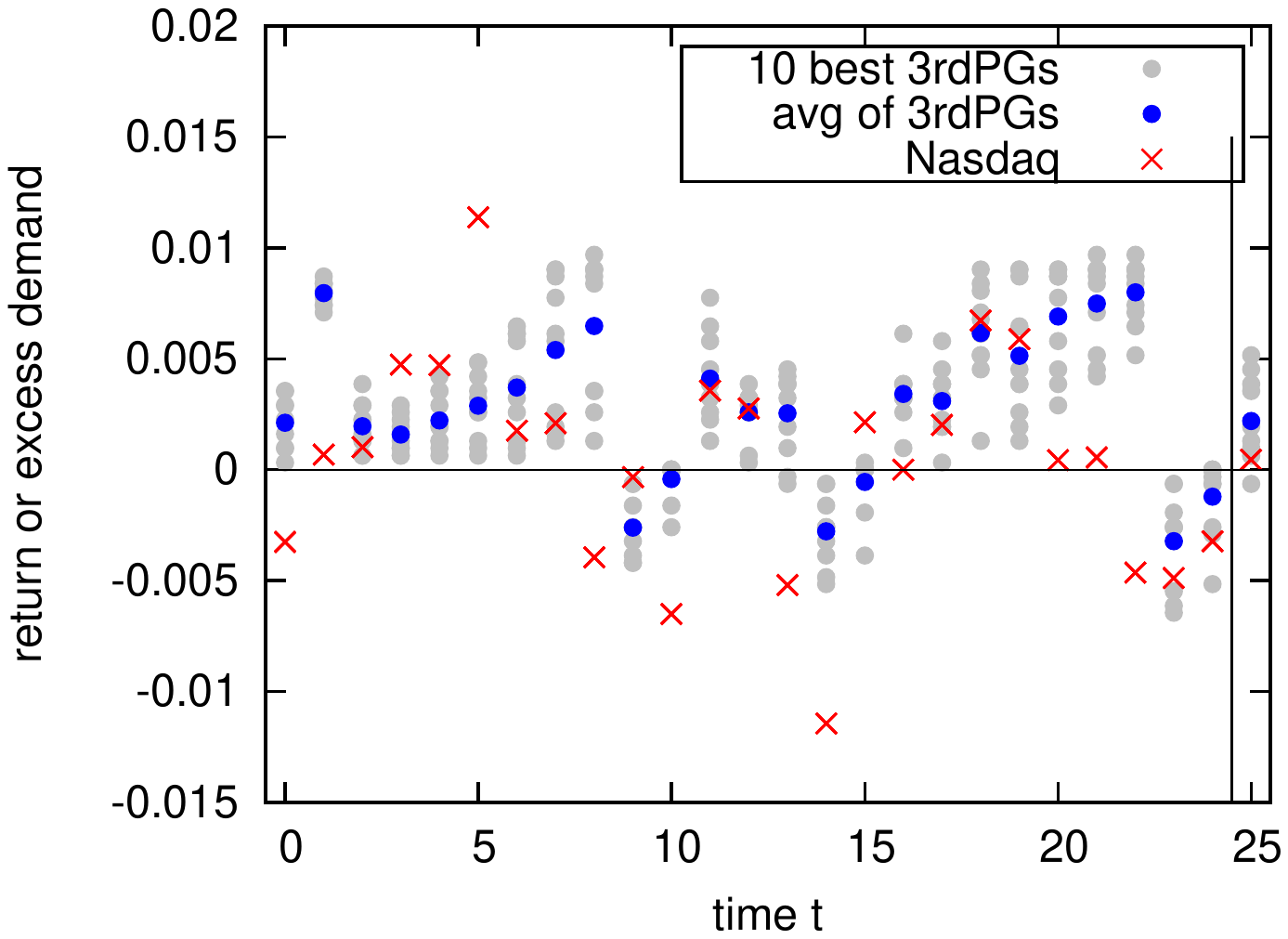}
{This is an illustration of the procedure which is repeated for each day out of the analyzed period of 606 days and for each of the 5 types of \acp{ABM} described in Subsection \ref{sec:abmTypes}. In the figure, the actual data (Nasdaq return illustrated as red crosses) and that generated by the reconstructed virtual stock market (the best 10 \acp{3PG} - in this sample consisting of \ac{GCMjG} agents - illustrated as gray dots and their average in blue) are plotted, whereby the vertical line separates the in-sample period during which their similarity is optimized (here 25 days) from the out-of-sample period (one-step prediction).}

Figure \ref{fig:3rdPGvsExtReturn} shows the excess demand obtained from the aggregate decisions of all the agents of one selected best \ac{3PG} for a given time window of the Nasdaq Composite index.
The data point to the right of the vertical line is the next-step, out-of-sample, prediction, whereas all points to the left of it belong to the in-sample period during which the \ac{3PG} has been trained on the external time series and has been optimized in terms of its \ac{ISD}. Every run of the \ac{GA} results in one best - according to its fitness - \ac{3PG} which then can be used to predict the next day return.

For each time window of the Nasdaq Composite index, we obtain the best \ac{3PG} for each of the five types of \acp{ABM} defined in Subsection \ref{sec:abmTypes}, using the above \ac{GA}.
This provides us with five different ``lenses'' to examine the Nasdaq Composite, that reveal its different characteristics.

\section{Validation by the statistical significance of the success rate of next-day prediction}
\label{sec:results}

In order to test the predictive value of the reverse engineered  \ac{3PG}
for each of the five \acp{ABM}, we report the success rate, that is, the fraction of days out-of-sample for which the predicted and realized returns have the same sign.

In order to assess the statistical significance of the obtained success rates, we compare
them to those of  1000 \textit{random strategies}, obtained by predicting with equal probability $1/2$ the rise or decline of each next-day market price. Using random strategies has been shown to provide the most robust estimations of the statistical significance of strategies in the presence of biases and trends \citep*{Daniel2009LookAhead}. In Table \ref{tab:accRates},
we report the $p$-value of the \acp{3PG} for each \ac{ABM}, calculated as the fraction of random strategies that perform better.

Table \ref{tab:accRates} reports the success rates and their corresponding p-values 
for each type of  \ac{ABM} averaged over 
(i) all 606 days, 
(ii) for the trending \footnote[1]{Bullish markets consisting of at least double the amount of days having a positive return than days having a negative return. In other words, on 2 out of 3 days the market goes up; vice-versa for bearish markets.} periods (202 days) and  
(iii) for the non-trending \footnote[2]{Not trending markets are composed of an equal amount of days on which the market is going up as going down.} periods (404 days). In the second column, the success rates are averaged over all parameter sets of the \ac{GA}. The third and fourth column report the minimum and maximum success rates over the parameter sets of the  \ac{GA}.

\begin{table}[tp]
\begin{center}
	\begin{tabular}{|l||c|c|c|}
		\hline
		\textbf{agent type} & \textbf{(p-val)	avg} & \textbf{min} & \textbf{max}  \\ \hline
		\multicolumn{4}{|c|}{All periods} \\ \hline
	  GCMG & (0.01)	0.55 & 0.51 & 0.60  \\
		GCMjG & (0.00)	0.57 & 0.54 & 0.60  \\
		delGCMjG & (0.00)	0.57 & 0.54 & 0.59  \\
		delGCMG & (0.02)	0.54 & 0.51 & 0.57  \\
		MixG & (0.00)	0.56 & 0.53 & 0.58  \\ \hline
		
		\multicolumn{4}{|c|}{Trending periods} \\ \hline
	  GCMG & (0.02)	0.57 & 0.54 & 0.63  \\
		GCMjG & (0.00)	0.66 & 0.63 & 0.68  \\
		delGCMjG & (0.00)	0.67 & 0.64 & 0.70  \\
		delGCMG & (0.01)	0.58 & 0.55 & 0.61  \\
		MixG & (0.00)	0.67 & 0.62 & 0.68  \\ \hline
		
		\multicolumn{4}{|c|}{Non-trending periods} \\ \hline
	  GCMG & (0.07)	0.53 & 0.49 & 0.58  \\
		GCMjG & (0.13)	0.53 & 0.49 & 0.55  \\
		delGCMjG & (0.24)	0.52 & 0.49 & 0.54  \\
		delGCMG & (0.15)	0.52 & 0.50 & 0.55  \\
		MixG & (0.33)	0.51 & 0.48 & 0.54  \\ \hline
	\end{tabular}
\end{center}
\caption{Success rates (average, minimum, and maximum) and their $p$-values (stated in parentheses) for each type of  \ac{ABM} cumulated over (i) all days, (ii) the trending periods, and (iii) the non-trending periods.
The trending periods cover 202 days from 1985-10-25 until 1986-03-20, and from 1984-01-05 until 1984-05-29.
The non-trending periods cover 404 days from 1976-05-10 until 1976-09-30, from 1984-04-05 until 1984-08-28, from 2002-06-20 until 2002-11-11, and from 2008-10-21 until 2009-03-17.}
\label{tab:accRates}
\end{table}

Over all days independently of the presence or absence of trends, 
the success rates of the reverse engineered \ac{3PG} are superior to all random strategies ($p < 0.001$) for the \ac{GCMjG}, \ac{delGCMjG}, and \ac{MixG}. For the \ac{GCMG} and \ac{delGCMG}, the results are still very significant with $p$-values given respectively by $0.01$ and $0.02$. 

Decomposing the 606 test days into trending and non-trending periods, we find that the success rates are very significant for the former periods and less so for the later periods.
The reverse engineering procedure is thus a good trend detection method. 

While it is expected that the  \ac{GCMjG}, \ac{delGCMjG}, and \ac{MixG} would perform in trending period due to their majority incentive, it is a priori quite surprising that the  \ac{GCMG} and \ac{delGCMG} also perform very significantly. We interpret this result as follows. First, the reverse engineering process applied with the \ac{GCMjG}, \ac{delGCMjG}, and \ac{MixG} selects the trend-following strategies which, when used
by a majority of agents, allow a good fit to the trend. Second, the fact that \ac{GCMG} and \ac{delGCMG} also perform well in trending periods implies that these trending periods are not just simple trends, but are decorated with cycles or alternating correction phases that the minority mechanism is able to pick up. 

In contrast, all \acp{ABM} show a strong drop in performance in the non-trending periods, with the best performing game being the \ac{GCMG}.
This later result can be rationalized by the minority incentive of this game, which is known to lead to oscillation prices resulting from the frustration inherent to the minority payoff \citep*{Marsili2001Market}.

\section{Conclusion}
\label{sec:conclusion}

In conclusion, we have shown that reverse engineering a real financial time series
with simple \acp{ABM} selected by using a genetic algorithm might be possible and provide
novel insight in the properties of financial time series. Notwithstanding the simplicity
(some would say ``naivety'') of the used \acp{ABM}, the aggregation of simple 
interactions at the micro level is sufficient to generate sophisticated structures at the macro level, which is probably the explanation for the good performance obtained in the validation step.

Finally, the method developed here is more generally applicable to the prediction of complex systems with an underlying multi-agent structure.

{\bf Acknowledgements}: We acknowledge financial support 
from the ETH Competence Center ``Coping with Crises in Complex 
Socio-Economic Systems" (CCSS) through ETH Research 
Grant CH1-01-08-2 and ETH Zurich Foundation.

\newpage
\appendix

\bibliographystyle{elsart-harv}

\end{document}